\documentstyle[prd,aps,preprint,psfig]{revtex}

\newcommand{\CL}{{\cal L}}

\newcommand{\CO}{{\cal O}}

\newcommand{\bear}{\begin{array}}  \newcommand{\eear}{\end{array}}
\newcommand{\bea}{\begin{eqnarray}}  \newcommand{\eea}{\end{eqnarray}}
\newcommand{\beq}{\begin{equation}}  \newcommand{\eeq}{\end{equation}}
\newcommand{\bef}{\begin{figure}}  \newcommand{\eef}{\end{figure}}
\newcommand{\bec}{\begin{center}}  \newcommand{\eec}{\end{center}}
\newcommand{\non}{\nonumber}  
\newcommand{\lmk}{\left(}  \newcommand{\rmk}{\right)}
\newcommand{\lkk}{\left[}  \newcommand{\rkk}{\right]}
\newcommand{\lhk}{\left \{ }  \newcommand{\rhk}{\right \} }

\newcommand{\bib}{\bibitem} 
\newcommand{\la}{\left\langle} \newcommand{\ra}{\right\rangle}

\newcommand{\gr}{g_{\rm R}}
\newcommand{\gi}{g_{\rm I}}

\def\IB#1#2#3{{\bf #1}, #2 (19#3)}

\def\IBID#1#2#3{{\it ibid}. {\bf #1}, #2 (19#3)}
\def\IBIDD#1#2#3{{\it ibid}. {\bf #1}, #2 (20#3)}

\def\APJ#1#2#3{Astrophys. J. {\bf #1}, #2 (19#3)}
\def\APJL#1#2#3{Astrophys. J. Lett. {\bf #1}, L#2 (19#3)}
\def\APJLL#1#2#3{Astrophys. J. Lett. {\bf #1}, L#2 (20#3)}

\def\CQG#1#2#3{Class. Quantum Grav. {\bf #1}, #2 (19#3)}

\def\MNRAS#1#2#3{Mon. Not. R. Astron. Soc. {\bf #1}, #2 (19#3)}
\def\MPLA#1#2#3{Mod. Phys. Lett. A {\bf #1}, #2 (19#3)}

\def\NATT#1#2#3{Nature (London) {\bf #1}, #2 (20#3)}
\def\NPB#1#2#3{Nucl. Phys. {\bf B#1}, #2 (19#3)}
\def\NPBB#1#2#3{Nucl. Phys. {\bf B#1}, #2 (20#3)}

\def\PLB#1#2#3{Phys. Lett. B {\bf #1}, #2 (19#3)}
\def\PLBB#1#2#3{Phys. Lett. B {\bf #1}, #2 (20#3)}
\def\PLBold#1#2#3{Phys. Lett. {\bf#1B}, #2 (19#3)}

\def\PRD#1#2#3{Phys. Rev. D {\bf #1}, #2 (19#3)}
\def\PRDD#1#2#3{Phys. Rev. D {\bf #1}, #2 (20#3)}

\def\PRLL#1#2#3{Phys. Rev. Lett. {\bf#1}, #2 (20#3)}

\def\PRT#1#2#3{Phys. Rep. {\bf#1}, #2 (19#3)}
\def\PTP#1#2#3{Prog. Theor. Phys. {\bf #1}, #2 (19#3)}

\begin{document}

\tighten
\draft
\title{Natural Double Inflation in Supergravity}
\author{Masahide Yamaguchi}
\address{Research Center for the Early Universe, University of Tokyo,
  Tokyo, 113-0033, Japan}

\date{\today}

\maketitle

\begin{abstract}
    We propose a natural double inflation model in supergravity. In
    this model, chaotic inflation first takes place by virtue of the
    Nambu-Goldstone-like shift symmetry, which guarantees the absence
    of the exponential factor in the potential for the inflaton field.
    During chaotic inflation, an initial value of the second inflation
    (new inflation) is set. In this model, the initial value of new
    inflation can be adequately far from the local maximum of the
    potential for new inflation due to the small linear term of the
    inflaton in the K\"ahler potential. Therefore, the primordial
    fluctuations within the present horizon scale may be attributed to
    both inflations; that is, the first chaotic inflation produces the
    primordial fluctuations on the large cosmological scales while the
    second new inflation on the smaller scales. The successive decay
    of the inflaton for new inflation leads to a reheating temperature
    low enough to avoid the overproduction of gravitinos in a wide
    range of the gravitino mass.
\end{abstract}



\section{Introduction}

\label{sec:int}

The fascinating feature of inflation is a generation of primordial
density fluctuations in addition to solving the flatness and the
horizon problems of the standard big bang cosmology. An single
inflation model generally predicts adiabatic fluctuations with a
nearly scale-invariant spectrum. It is, however, pointed out that a
standard cold dark matter model with a nearly scale-invariant spectrum
cannot account for several observational results. The predicted
primordial fluctuations have too much powers on small scales after
they are normalized by the cosmic microwave background explorer (COBE)
data. The power spectrum derived from the APM redshift survey has a
break around the scale $k \sim 0.05 h$ Mpc$^{-1}$ though there is
large uncertainty \cite{APM}. Furthermore, while the recent
observations of anisotropies of the cosmic microwave background (CMB)
by the BOOMERANG experiment \cite{BOOMERANG} and the MAXIMA experiment
\cite{MAXIMA} confirmed a flat universe with an inflationary scenario,
they found the peculiar feature, that is, a relatively low second
acoustic peak. Thus, the demand for the primordial fluctuations with a
non-trivial shape is increasing.

Such primordial fluctuations are realized in a double inflation model. 
In fact, double inflation models have been considered to reconcile the
predicted spectra with the observations
\cite{double,SUSYdouble1,SUSYdouble2}. Many of them were discussed in
a simple toy model with two massive scalar fields. Though the model
predicts an interesting nontrivial spectrum, a more realistic double
inflation model should be constructed on the basis of the particle
physics theory. Supersymmetry (SUSY) is one the most powerful
extensions of the standard model of particle physics, which is also
indispensable to constructing inflation models. Because SUSY
guarantees the flatness of the inflaton and gives a natural solution
to the hierarchy problem between the inflationary scale and the
electroweak scale. Therefore, it is very important to construct double
inflation models in the context of SUSY, especially, its local
version, supergravity (SUGRA) \cite{SUSY,LR}. Until now, a few double
inflation models in SUSY or SUGRA have been proposed
\cite{SUSYdouble1,SUSYdouble2,SUSYPBF}. First of all, double hybrid
inflation models in SUSY \cite{PT} or SUGRA \cite{LT} have been
proposed in order to solve the initial condition problem of hybrid
inflation \cite{inithybrid}. Later, it is shown that they can produce
the primordial fluctuations with the non-trivial feature as observed
\cite{SUSYdouble1}. But, generally speaking, the reheating temperature
after hybrid inflation is so high that it causes the overproduction of
gravitinos. On the other hand, Izawa {\it et al.} \cite{IKY} proposed
preinflation as a solution to the initial condition problem of new
inflation, which straightforwardly predicts the low reheating
temperature \cite{initnew}. If hybrid inflation is taken as
preinflation, double inflation is realized through the cross term of
the superpotential. Later, it has been discussed in the context of
large scale structure, CMB \cite{SUSYdouble2}, and primordial black
holes (PBHs) formation \cite{SUSYPBF}.

In all of the earlier double inflation models in SUSY or SUGRA, hybrid
inflation is adopted as the first inflation responsible for the COBE
scale. However, such hybrid inflation is known to suffer from the
severe initial condition problem \cite{inithybrid}. Hence, another
hybrid inflation is required, which takes place near the Planck scale
and the produced fluctuations cannot be observed
\cite{PT,LT}.\footnote{Exactly speaking, even if one considers such
pre-hybrid inflation, it suffers from the so-called flatness
(longevity) problem, that is, why the universe lives beyond the Planck
time (though it is milder than the original one) as long as it starts
below the Planck scale.} After all, triple inflation is required. On
the other hand, chaotic inflation \cite{chaoinf} is free from any
initial condition problems including the flatness problem because it
starts around the Planck scale. Nonetheless, one of the reasons why
hybrid inflation has often been considered as an inflation model in
SUGRA lies in difficulty realizing chaotic inflation in SUGRA. The
minimal SUGRA potential for scalar fields has an exponential factor of
the form $\exp(|\phi|^2/M_G^2)$, which prevents any scalar field
$\phi$ from taking a value much larger than the reduced Planck scale
$M_G\simeq 2.4\times 10^{18}$ GeV. Several chaotic inflation models
have been proposed so far by use of the functional degrees of freedom
in SUGRA \cite{GL,MSY2}. Rather specific K\"ahler potentials were
adopted without symmetry reasons so that we need the fine tuning.
Recently, Kawasaki {\it et al.} \cite{KYY} proposed a natural model of
chaotic inflation in SUGRA, where the K\"ahler potential is restricted
by the Nambu-Goldstone-like shift symmetry.

In this paper, we propose a natural double inflation model in SUGRA by
use of the Nambu-Goldstone-like shift symmetry. In this model, first
of all, chaotic inflation takes place. During chaotic inflation, an
initial value of the second inflation (new inflation) is set due to
the supergravity effects. A similar model has already been proposed by
Yokoyama and Yamaguchi \cite{YY}(see also Ref. \cite{Yokoyama}).
But, in the model, the initial value of new inflation is so close to
the local maximum of the potential for new inflation, which causes the
universe to enter a self-regenerating stage \cite{eternal,sr}.  Hence,
the primordial fluctuations responsible for the observable scale are
produced only during the last inflation (new inflation) so that their
spectra are only tilted scale-free ones. Also, you should notice that
even if the chaotic inflation proposed in Ref. \cite{KYY} is taken as
preinflation in Ref. \cite{IKY}, the same situation occurs, that is,
the second inflation becomes eternal inflation because the
superpotential in Ref. \cite{KYY} vanishes during chaotic inflation.
On the other hand, in our model, the initial value of new inflation
can be adequately far from the local maximum of the potential due to
the small linear term of the inflaton in the K\"ahler potential.
Therefore, the primordial fluctuations responsible for the observable
universe may be attributed to both inflations, that is, the first
chaotic inflation produces the primordial fluctuations on the large
cosmological scales while the second new inflation on the smaller
scales. The fact that the last inflation is new inflation is favorable
because it straightforwardly predicts sufficiently low reheating
temperature to avoid the overproduction of gravitinos. Furthermore,
our model is simple in that two inflatons belong to the same
supermultiplet, namely, one direction of a complex scalar field drives
chaotic inflation while another drives new inflation. Also, our model
is natural for two reasons. The form of K\"ahler potential is
completely determined by a symmetry, that is, the Nambu-Goldstone-like
shift symmetry. Next, though we need the introduction of small
breaking parameters, the smallness of parameters is justified also by
symmetries. That is, the zero limit of small parameters recovers
symmetries, which is natural in the 't Hooft's sense \cite{tHooft}.

In the next section, we present our double inflation model in
supergravity. In Sec. III, the dynamics is investigated and the
primordial density fluctuations are estimated. In the final section,
we give discussions and conclusion.

\section{Model}

\label{sec:model}

We introduce an inflaton chiral superfield $\Phi(x,\theta)$ and a
spurion superfield $\Xi$. We assume that the model is invariant under
the following Nambu-Goldstone like shift symmetry \cite{KYY}:
\bea
  \Phi &\rightarrow& \Phi + i~C M_{G}, \non \\
  \Xi  &\rightarrow& \lmk\frac{\Phi}{\Phi + i~C M_{G}}\rmk^{2} \Xi,
  \label{eq:shift}
\eea
where $C$ is a dimensionless real constant and $M_{G}$ is the reduced
Planck scale. That is, the combination $\Xi\Phi^{2}$ is invariant
under the shift symmetry. Then, the K\"ahler potential is a function
of $\Phi + \Phi^{\ast}$, i.e. $K(\Phi,\Phi^{\ast}) = K(\Phi +
\Phi^{\ast})$, which allows the imaginary part of the scalar
components of $\Phi$ to take a value much larger than the
gravitational scale. Later we set $M_{G}$ to be unity.

Next, let us discuss the form of the superpotential. We assume that
the superpotential is also invariant under the $U(1)_{\rm R}$
symmetry, which prohibits a constant term in the superpotential. Then,
the earlier K\"ahler potential is invariant only if the R charge of
$\Phi$ is zero, which compels us to introduce another supermultiplet
$X(x,\theta)$ with its R charge equal to two. The general
superpotential invariant under the shift and the $U(1)_{\rm R}$
symmetries is given by
\bea
  W = v X \lkk 1 + \alpha_{2}(\Xi\Phi^{2})^{2} + \cdots \rkk
      - X \lkk \Xi\Phi^{2} + \alpha_{3}(\Xi\Phi^{2})^{3} + \cdots
      \rkk,
  \label{eq:superpotential}
\eea

where $v$, $\alpha_{i}$ are complex constants and we have assumed the
R charge of $\Xi$ vanishes. Generally speaking, among all complex
constants, only one constant can become real by use of the phase
rotation of the $X$ field. Later we set $v$ to be real. The shift
symmetry is softly broken by inserting the vacuum value into the
spurion field, $\la \Xi \ra = \lambda$. The parameter $\lambda$ is
fixed at an value whose absolute magnitude is much smaller than unity,
representing the magnitude of breaking of the shift symmetry [Eq.
(\ref{eq:shift})]. As long as $|\Phi| \ll |\lambda|^{-1/2}$, higher
order terms with $\alpha_{i}$ of the order of unity become irrelevant
for the dynamics of the chaotic inflation. Thus, we can safely neglect
them in the following discussion. As shown later, for successful
inflation, the constant $v$ must be at most of the order of
$|\lambda|$, which is much smaller than unity. Since the constant $v$
is of the order of unity in general, we must invoke the mechanism to
suppress the constant $v$. For the purpose, we introduce the $Z_{2}$
symmetry, under which the $\Phi$, $X$, and $\Xi$ fields are
odd.\footnote{The fact that the $Z_{2}$ charge of $\Phi$ is odd is
essential for this model because it allows the small linear term of
$\Phi + \Phi^{\ast}$ in the K\"ahler potential so that the initial
value of new inflation can appropriately deviate from the local
maximum of the potential for new inflation.}\footnote{Note that the
spurion field $\Xi$ in fact breaks both the shift symmetry and the
$Z_{2}$ symmetry at once. So, we expect that the magnitudes of the
breaking of both the $Z_{2}$ and the shift symmetries are of the same
order, that is, $|g| = \CO(1)$.}(See Table I in which charges for
superfields are shown.) Then, the smallness of the constant $v$ is
associated with the small breaking of the $Z_{2}$ symmetry. That is,
we introduce a spurion field $\Pi$ with the odd $Z_{2}$ charge and the
zero R charge, whose vacuum value $\la \Pi \ra = v$ softly breaks the
$Z_{2}$ symmetry. You should notice that though the above
superpotential is not invariant under the shift and the $Z_{2}$
symmetries, the model is completely natural in 't Hooft's sense
\cite{tHooft} because we have enhanced symmetries in the limits
$\lambda \rightarrow 0$ and $v \rightarrow 0$. In the following
analysis, we use the superpotential given by
\bea
  W &\simeq& vX - \lambda X\Phi^{2}, \\
    &=& vX(1 - g\Phi^{2}),
\eea
with $g \equiv \lambda / v$. 

The K\"ahler potential invariant under the shift and the $U(1)_{\rm
R}$ symmetries is given by
\beq
  K = v_{2}(\Phi + \Phi^{\ast}) 
     + \frac12 (\Phi + \Phi^{\ast})^{2} 
     + XX^{\ast}  + \cdots.
  \label{eq:kahler}
\eeq
Here $v_{2} \sim v$ is a real constant representing the breaking
effect of the $Z_{2}$ symmetry. The term $v_{3}\lambda_{3}\Phi^{2} +
v_{3}^{\ast}\lambda_{3}^{\ast}{\Phi^{\ast}}^{2}$ may appear, where
$v_{3}$ and $\lambda_{3}$ are complex constants representing the
breaking of the $Z_{2}$ and the shift symmetries ($|v_{3}| \sim v$ and
$|\lambda_{3}| \sim |\lambda_{4}| \sim |\lambda|$). But, these terms
are extremely small so that we can safely omit them in the K\"ahler
potential [Eq. (\ref{eq:kahler})]. A constant term is also omitted because
it only changes the overall factor of the potential, whose effect can
be renormalized into the constants $\lambda$ and $\delta_{1}$. Here
and hereafter, we denote the scalar components of the supermultiplets
by the same symbols as the corresponding supermultiplets.

\section{Dynamics of double inflation and primordial density fluctuations}

\label{sec:dynamics}

Neglecting higher order terms and a constant term in the K\"ahler
potential, the Lagrangian density $L(\Phi,X)$ for the scalar fields
$\Phi$ and $X$ is given by
\beq
  L(\Phi,X) = \partial_{\mu}\Phi\partial^{\mu}\Phi^{\ast} 
  + \partial_{\mu}X\partial^{\mu}X^{\ast}
         -V(\Phi,X),
\eeq
where the scalar potential $V$ of the chiral superfields $X(x,\theta)$
and $\Phi(x,\theta)$ is given by
\beq
  V = v^{2} e^{K} \lkk
      \left|1 - g\Phi^{2}\right|^{2}(1-|X|^{2}+|X|^{4}) 
       + |X|^{2} \left
         |-2g\Phi + (v_{2}+\Phi+\Phi^{\ast})(1-g\Phi^{2})
                 \right|^{2}
                  \rkk.
\eeq
Decomposing the scalar field $\Phi$ and the complex constant $g$ into
real and imaginary components
\bea
  \Phi &=& \frac{1}{\sqrt{2}} (\varphi + i \chi), \\
  g &=& \gr + i \gi,
\eea
the Lagrangian density $L(\varphi,\chi,X)$ is written as
\beq
  L(\varphi,\chi,X) = 
              \frac{1}{2}\partial_{\mu}\varphi\partial^{\mu}\varphi 
              + \frac{1}{2}\partial_{\mu}\chi\partial^{\mu}\chi 
              + \partial_{\mu}X\partial^{\mu}X^{*}
              -V(\varphi,\chi,X),
\eeq
with the potential $V(\varphi,\chi,X)$ given by
\bea
  V(\varphi,\chi,X)
    &=& v^{2} e^{-\frac{v_{2}^{2}}{2}}
           \exp \lkk \lmk \varphi + \frac{v_{2}}{\sqrt{2}} \rmk^{2} 
                     + |X|^{2}
                \rkk \non \\ 
    && \hspace{-2.0cm} \times
         \lhk~\lkk 
              1 - \gr (\varphi^{2} - \chi^{2}) + 2 \gi\varphi\chi 
             + \frac14~(\gr^{2} + \gi^{2})~(\varphi^{2} + \chi^{2})^{2}
              \rkk
             (1-|X|^{2}+|X|^{4}) 
         \right. \non \\ 
    && \hspace{-1.5cm}
             +~|X|^{2} 
              \lkk~
                2~(\gr^{2} + \gi^{2})~(\varphi^{2}+\chi^{2})
              \right. \non \\
    && \hspace{-0.0cm}
                -~(v_{2}+\sqrt{2}\varphi) \lhk
                  \sqrt{2}~(\gr\varphi - \gi\chi)
                    \lkk~2 - \gr (\varphi^{2} - \chi^{2}) 
                         + 2 \gi \varphi\chi
                    ~\rkk \right. \non \\
    && \hspace{+2.5cm} \left.
                  - \sqrt{2}~(\gr\chi + \gi\varphi)
                    \lkk~\gi (\varphi^{2} - \chi^{2}) 
                         + 2 \gr \varphi\chi
                    ~\rkk~\rhk \non \\
    && \hspace{-0.0cm} \left. \left.    
                 +~(v_{2}+\sqrt{2}\varphi)^{2} 
                    \lhk
                      1 - \gr (\varphi^{2} - \chi^{2}) + 2 \gi\varphi\chi 
                    + \frac14~(\gr^{2} + \gi^{2}) 
                    ~(\varphi^{2} + \chi^{2})^{2}
                    \rhk
               ~\rkk
                      ~\rhk.
\eea

\subsection{chaotic inflation}

Though the potential is very complicated, the dynamics is not so.
While $\varphi, |X| \lesssim \CO(1)$ due to the factor $e^{v_{2}
\varphi + \varphi^{2} + |X|^{2}}$ ($v_{2} \ll 1$), $\chi$ can take a
value much larger than unity without costing exponentially large
potential energy. Then, the scalar potential is approximated as
\beq
  V \simeq |\lambda|^{2}
                   \lmk \frac{\chi^{4}}{4}      
                        + 2 \chi^{2} |X|^{2}
                   \rmk,
  \label{eq:twopot}
\eeq
with $|\lambda|^{2} = (\gr^{2} + \gi^{2}) v^{2}$. Thus, the term
proportional to $\chi^{4}$ becomes dominant, which leads to chaotic
inflation starting around the Planck epoch. Using the slow-roll
approximations, we obtain the $e$-fold number $N_{c}$,
\beq
  N_{c} \simeq \frac{\chi_{N_{c}}^{2}}{8}.
\eeq
During chaotic inflation, the potential minimum for $\varphi$,
$\varphi_{\rm min}$, is given by
\beq
  \varphi_{\rm min} \simeq - v_{2}/\sqrt{2} 
                       - \frac{g_{I}}{g_{R}^{2}+g_{I}^{2}}
                          \frac{4}{\chi^{3}},
\eeq
and the mass squared of $\varphi$, $m_{\varphi}^{2}$, becomes
\beq
  m_{\varphi}^{2} \simeq \frac{|\lambda|^{2}}{2} \chi^{4}
                  \simeq 6H^{2} \gg \frac94 H^{2},
                  ~~~~~~H^{2} \simeq \frac{|\lambda|^{2}}{12}\chi^{4},
\eeq
where $H$ is the hubble parameter at that time. Hence, $\varphi$
oscillates rapidly around the minimum $\varphi_{\rm min}$ with its
amplitude damping in proportion to $a^{-3/2}$ ($a$ : the scale
factor of the universe). Thus, at the end of chaotic inflation,
$\varphi$ settles down to the minimum $\varphi_{\rm min}$.

On the other hand, the mass squared of $X$, $m_{X}^{2}$, is dominated
by
\beq
  m_{X}^{2} \simeq 2 |\lambda|^{2} \chi^{2} \simeq \frac{24}{\chi^2}H^2,   
  \label{mhratio}
\eeq
which is much smaller than the hubble parameter squared in the early
stage of chaotic inflation so that $X$ also slow-rolls towards the
origin. Later we set $X$ to be real and positive making use of the
freedom of the phase choice. In this regime, the classical equations
of motion for the $X$ and $\chi$ fields are given by
\bea
  3H &\dot{X}& \simeq - m_{X}^{2} X,  \label{Xeq} \\
  3H &\dot{\chi}& \simeq - |\lambda|^{2} \chi^{3}, \label{chieq}
\eea
which yield
\beq
  X \propto \chi^{2}.   \label{propto}
\eeq
This relation holds actually if and only if quantum fluctuations are
unimportant for both $\chi$ and $X$. However, following the same
procedure done in Refs. \cite{KYY,YY}, we can easily show that for $X$,
quantum fluctuations are smaller than the classical value and $X$ is
much smaller than unity throughout chaotic inflation.

We investigate the density fluctuations produced by this chaotic
inflation. As shown above, there are the two effectively massless
fields, $\chi$ and $X$. Using Eq. (\ref{eq:twopot}) and adequate
approximations, the metric perturbation in the longitudinal gauge
$\Phi_A$ can be estimated as \cite{PS}
\bea
  \Phi_A &=& - \frac{\dot{H}}{H^{2}} C_{1}
               - 16 \frac{X^{2}}{\chi^{2}} C_{3}, \non \\
  C_{1} &=& H \frac{\delta\chi}{\dot{\chi}}, \non \\
  C_{3} &=& H \lmk \frac{\delta\chi}{\dot{\chi}} 
                  - \frac{\delta X}{\dot{X}}
              \rmk 
            \frac{2}{\chi^{2}},
\eea
where the dot represents the time derivative, the term proportional to
$C_{1}$ corresponds to the growing adiabatic mode, and the term
proportional to $C_{3}$ the nondecaying isocurvature mode. You should
notice that only $\chi$ contributes to the growing adiabatic
fluctuations.  Then, the amplitude of curvature perturbation $\Phi_A$
on the comoving horizon scale at $\chi=\chi_{N_{c}}$ is estimated by
the standard one-field formula as
\beq
  \Phi_A(N_{c}) \simeq  \frac{f}{2\sqrt{3}\pi} \frac{V^{3/2}}{V'}
            \simeq  \frac{f}{2\sqrt{3}\pi}
                       \frac{|\lambda|\chi_{N_{c}}^{3}}{8}, 
\eeq
where $f=3/5~(2/3)$ in the matter (radiation) domination. Later we
consider the case where the comoving scale corresponding to the COBE
scale exits the hubble horizon during chaotic inflation. Defining
$N_{\rm COBE}$ as the e-fold number corresponding to the COBE scale, the
COBE normalization requires $\Phi_A(N_{\rm COBE}) \simeq 3\times 10^{-5}$
\cite{COBE}, which yields
\beq
    |\lambda| \simeq 4.2 \times 10^{-3} \chi_{N_{\rm COBE}}^{-3}. 
\eeq
The spectral index $n_{s}$ is given by
\beq
  n_{s} \simeq 1 - \frac{3}{N_{\rm COBE}}.
\eeq

\subsection{New inflation}

As $\chi$ becomes of the order of unity, the dynamics becomes a little
complicated because the term with $v^{2} g_{R} \chi^{2}$ or the
constant term $v^{2}$ may become dominant. Depending on the parameters
$g_{R}$ and $g_{I}$, we have a break or no break between chaotic and
new inflation.\footnote{Exactly speaking, the constant term may become
dominant (small hybrid inflation) before new inflation starts.}

In order to investigate when new inflation starts, we define the
following two fields $\varphi'$ and $\chi'$,
\beq
  \left(
    \begin{array}{c}
      \varphi' \\
      \chi'
    \end{array}
  \right)
  = 
  \left(
    \begin{array}{cc}
       \cos{\theta} & -\sin{\theta} \\
       \sin{\theta} & \cos{\theta} 
    \end{array}
   \right)
  \left(
    \begin{array}{c}
      \varphi \\
      \chi
    \end{array}
  \right),
\eeq
where $\theta$ is a constant characterized by
\beq
  \cos{2\theta} = \frac{g_{R}}{\sqrt{g_{R}^{2} + g_{I}^{2}}} ,
  \qquad
  \sin{2\theta} = \frac{g_{I}}{\sqrt{g_{R}^{2} + g_{I}^{2}}} .
\eeq
Then, the potential with $X \simeq 0$ is rewritten as
\bea
  V(\varphi',\chi',X \simeq 0)
    &\simeq& v^{2} e^{-\frac{v_{2}^{2}}{2}}
           \exp \lkk \lmk \varphi' \cos{\theta}
                        + \chi' \sin{\theta}
                        + \frac{v_{2}}{\sqrt{2}} 
                     \rmk^{2} 
                \rkk \non \\ 
    && \hspace{-2.0cm} \times
         \lkk 
             \lmk 
               1 - \frac{|g|}{2} \varphi'^{2} 
             \rmk^{2}
            + \chi'^{2}
             \lmk
               |g| + \frac{|g|^{2}}{2} \varphi'^{2}
                   + \frac{|g|^{2}}{2} \chi'^{2}
             \rmk
         \rkk.  
\eea
We find that the global minima are given by $\varphi'^{2} = 2/|g|$ and
$\chi' = 0$. New inflation can take place if $|g| \gtrsim
\cos^{2}{\theta}$, that is, $|\lambda| \gtrsim v \cos^{2}{\theta}$.
Since the mass squared for $\varphi'$, $m_{\varphi'}^{2}$, is given by
\beq
  m_{\varphi'}^{2} 
      \simeq - |g| + \frac12 |g|^{2} \chi'^{2}
             + \cos^{2}{\theta} 
                 \lmk           
                   1 + |g| \chi'^{2} + \frac14 |g|^{2} \chi'^{4}
                 \rmk,
\eeq
new inflation begins when $\chi' \simeq \chi'_{\rm crit}$, which is
defined as
\beq
  \chi'_{\rm crit} \equiv  \frac{2}{|g|} 
                        \sqrt{\frac{|g| - \cos^{2}{\theta}}
                                   {|g| + 2 \cos^{2}{\theta}}}.
\eeq

Hereafter, we set $g_{I} = 0 (\cos{\theta} = 1, \sin{\theta} = 0)$ for
simplicity. In this case, for $\chi \simeq 0$ and $X \ll 1$, the
potential is approximated as
\beq
  V(\varphi,\chi \simeq 0,X \ll 1) \sim
    v^{2} \lkk 1 - (g_{R} - 1) \tilde\varphi^{2} 
                 + 2 (g_{R} - 1)^{2} \tilde\varphi^{2}|X|^{2}
                 + \cdots
          \rkk,
\eeq
with $\tilde\varphi = \varphi - \varphi_{\rm max}$ and $\varphi_{\rm max}
\equiv v_{2}/(\sqrt{2}(g_{R}-1))$. If $g_{R} \gtrsim 1$,\footnote{In
fact, for $g_{R} \gtrsim 1$, we can show that there is no local
minimum between the local maximum $\varphi_{\rm max}$ and the global
minima.} $\varphi$ slow-rolls down toward the vacuum expectation value
$\eta = \sqrt{2/g_{R}}$ and new inflation takes place.

In our model, $\varphi$ stays at $\varphi_{\rm min}$ until new inflation
starts. Thus, the initial value of $\tilde\varphi$,
$\tilde\varphi_{i}$, for new inflation is given by
\beq
  \tilde\varphi_{i} = - \frac{v_{2}}{\sqrt{2}}
                         \frac{g_{R}}{g_{R}-1}.
\eeq
The e-fold number $N_{n}$ is estimated as
\beq
  N_{n} \simeq \frac{1}{2(g_{R}-1)}
                \ln \left| \frac{\sqrt{2}}{v_{2}}
                           \frac{g_{R}-1}{g_{R}} \right|.
\eeq

In the new inflation regime, both $\varphi$ and $X$ acquire large
quantum fluctuations because they are effectively massless fields.
However, following the same procedure as done in the chaotic inflation
regime, it is shown that only $\varphi$ contributes to the adiabatic
fluctuations. The amplitude of curvature perturbation $\Phi_A$ on the
comoving horizon scale at $\tilde\varphi=\tilde\varphi_{N_{n}}$ is
given by
\beq
  \Phi_A(N_{n}) \simeq \frac{f}{2\sqrt{3}\pi} 
  \frac{v}{ 2 (g_{R} - 1) \tilde\varphi_{N_{n}}}, 
\eeq
where $f=3/5~(2/3)$ in the matter (radiation) domination. 
The spectral index $n_s$ of the density fluctuations is given by
\beq
    n_s \simeq 1 - 4 (g_{R} - 1).
\eeq

Now, let us comment on the amplitude of the density fluctuations in
the case with a break. Density fluctuations with comoving wave number
corresponding to the horizon scale around the beginning of new
inflation (if any, hybrid inflation) are induced during both chaotic
and new inflation. However, following the procedure as done in Refs.
\cite{SUSYdouble2,SUSYPBF}, we can show that the density fluctuations
produced during chaotic inflation are a little less than newly induced
fluctuations at the beginning of new inflation [$\simeq
v/(2\pi\sqrt{3}$)]. Furthermore, the fluctuations produced during
chaotic inflation are more suppressed for smaller wavelength. Thus, we
assume that the fluctuations of $\varphi$ induced during chaotic
inflation can be neglected when we estimate the amplitude of the
density fluctuations during new inflation.

After new inflation, $\varphi$ oscillates around the minimum $\varphi
= \eta$ so that the universe is dominated by a coherent oscillation of
the scalar field $\sigma \equiv \varphi -\eta$. Expanding the
exponential factor $e^{v_{2}\varphi + \varphi^2} $ in $e^K$
\beq
  e^{v_{2}\varphi + \varphi^2} = e^{\eta^2}(1+2\eta\sigma+\cdots ),
\eeq
we find that $\sigma$ has gravitationally suppressed linear
interactions with all scalar and spinor fields including minimal
supersymmetric standard model (MSSM) particles. For example, let us
consider the Yukawa superpotential $W = y_{i}D_{i}HS_{i}$ in MSSM,
where $D_{i},S_{i}$ are doublet (singlet) superfields, $H$ represents
Higgs superfields, and $y_{i}$ are Yukawa coupling constants. Then,
the interaction Lagrangian is given by
\beq
  \CL_{\rm int} \sim 
     y_{i}^{2} \eta \sigma D_{i}^{2} S_{i}^{2} + \dots,
\eeq
which leads to the decay width $\Gamma$ given by 
\beq
  \Gamma \sim \Sigma_{i} y_{i}^4 \eta^2 m_{\sigma}^3.
  \label{eq:gamma}
\eeq
Here $m_{\sigma} \simeq 2\sqrt{g_{R}}e^{\sqrt{2/g_{R}}}v$ is the mass
of $\varphi$ at the vacuum expectation value $\eta = \sqrt{2/g_{R}}$.
Then, the reheating temperature $T_{R}$ is given by
\beq
  T_{R} \sim 0.1 \bar{y} \eta m_{\varphi}^{3/2},
\eeq
with $\bar{y}=\sqrt{\Sigma_{i} y_{i}^4}$. Taking $\bar{y} \sim 1$, the
reheating temperature $T_{R}$ is given by
\beq
  T_{R} \sim v^{3/2} \lesssim |\lambda|^{3/2} 
         \sim 10^{-4} \chi_{N_{\rm COBE}}^{-9/2}. 
\eeq
For $N_{\rm COBE} = \CO(10)$, $T_{R}$ is low enough to avoid the
overproduction of gravitinos in a wide range of the gravitino mass
\cite{Ellis}.

\section{Discussion and conclusions}

\label{sec:con}

In this paper, we propose a natural double inflation model in SUGRA.
By use of the shift symmetry, first of all, chaotic inflation can take
place, which has no initial condition problem. Later, new inflation
takes place, which straightforwardly leads to the low reheating
temperature enough to avoid the overproduction of gravitinos. In our
model, the initial value of new inflation is set during chaotic
inflation and is appropriately far from the local maximum of the
potential for new inflation so that both inflations are responsible
for the primordial fluctuations on the observable scale. In the
forthcoming paper, we will investigate the density fluctuations
produced during both inflations in detail and compare them with recent
observations. Moreover, we will discuss the PBHs formation. Since the
second inflation is new inflation in our model, the produced density
fluctuations become of the order of unity due to the peculiar property
of new inflation. PBHs may be identified with the massive compact halo
objects (MACHOs) or be responsible for antiproton fluxes observed by
the BESS experiments.

\subsection*{ACKNOWLEDGMENTS}

M.Y. is grateful to M. Kawasaki and J. Yokoyama for many useful
discussions and comments. M.Y. is partially supported by the Japanese
Society for the Promotion of Science.

\begin{table}[t]
  \begin{center}
    \begin{tabular}{| c | c | c | c | c |}
                   & $\Phi$ & $X$ & $\Xi$ & $\Pi$ \\
        \hline
        $Q_R$      & 0      & 2   & 0     & 0 \\ 
        \hline 
        $Z_{2}$    & $-$    & $-$ & $-$   & $-$   
    \end{tabular}
    \caption{The charges of the $U(1)_{\rm R} \times Z_{2}$ symmetries
    for the various supermultiplets.}
    \label{tab:charges}
  \end{center}
\end{table}

\end{document}